# Non-intrusive Load Monitoring via Multi-label Sparse Representation based Classification

Shikha Singh and Angshul Majumdar

*Abstract*— This work follows the approach of multi-label classification for non-intrusive load monitoring (NILM). We modify the popular sparse representation based classification (SRC) approach (developed for single label classification) to solve multi-label classification problems. Results on benchmark REDD and Pecan Street dataset shows significant improvement over state-of-the-art techniques with small volume of training data.

## I. Introduction

IN non-intrusive load monitoring (NILM) the technical goal is to estimate the power consumption of different appliances given the aggregate smart-meter readings [1]. The broader social objective is to feedback this information to the household so that they can reduce power consumption and thereby save energy.

Over time, various approaches have been proposed to address this problem ranging from combinatorial optimization [2] and stochastic finite state machines [3] to modern deep learning based techniques [4]. A slightly dated review on this topic is available in [5, 6].

Strictly speaking traditional NILM is not fully non-intrusive; the data collection for the training stage is highly intrusive requiring installation of sensors at the plug level to record the consumption of individual appliances over months. This training data is used to train a model, which is then used in the operational / testing stage to disaggregate the load; the operational stage is non-intrusive.

Owing to the high cost of data collection, financial and privacy-wise, large scale roll-out of NILM as a service has not be achieved. There is a need to make the process completely non-intrusive (at least as far as sensing is concerned). The recent multi-label classification based framework for NILM is showing promise [7] in this direction. In this approach, the actual power consumption of the appliances is not required, only the ON/OFF state of the device needs to be recorded. This can be done by recording logs from individual households for residential buildings and building managers for commercial ones. Such an approach reduces both instrumentation costs and mitigates privacy concerns in one go.

In the multi-label classification based approach the states of the appliances are the class label. The recorded smart-meter readings serve as the input sample. Since multiple appliances can be ON at the same time, it turns out to be a multi-label classification problem; the interested reader should peruse [7] for details. In [7] a thorough comparison of traditional multi-label classification algorithms like multi-label K nearest neighbour (MLKNN) and random K label-sets (RaKEL) for NILM have been performed. More modern approaches are based on multi-label deep learning [8] and multi-label graph learning [9] for NILM; these form the state-of-the-art in this area.

In this work we propose a new approach to multi-label classification based on the sparse representation based classification (SRC) approach [10]. The technique was originally developed for computer vision, but has been widely used in various domains since then; the paper has 8000+ citations. The main advantage of SRC over other approaches is its ability to infer from very few samples. This is a critical criterion for NILM – the smaller the training data required the better it is. Compare two scenarios for a domestic household – training data logged for 7-8 months versus data logged for one month. In the first one, it is likely that the household will refuse to participate for two main reasons:
1. logging the data is tedious
2. household cannot go for vacation in this duration

These issues are likely not arise for the second scenario where they have to log the data for just a month. This is the reason, we are emphasizing on a mechanism that can infer from small volume of training data.

Originally SRC was developed for single label classification, i.e. for problems where the input samples corresponded to only one class label. Here we show how SRC can be easily extended to handle multi-label classification problems.

## II. Proposed Approach

SRC assumes that the test sample ($v_k$) can be represented as a linear combination of training samples from the correct ($k^{th}$) class. This is represented as follows,

$$v_{test} = \alpha_{k,1}v_{k,1} + \alpha_{k,2}v_{k,2} + ... + \alpha_{k,n_k}v_{k,n_k} + \varepsilon_k = V_k\alpha_k + \varepsilon_k \quad (1)$$

Here $v_{k,i}$ represents training samples for the $k^{th}$ class and $\alpha_{k,i}$ the corresponding linear weights; $V_k$ is formed by stacking the $v_{k,i}$'s as columns and $\alpha_k$ is formed by $\alpha_{k,i}$'s stacked as a vector. The error $\varepsilon_k$ is assumed to be Normally distributed.

However, the correct class is not known, therefore a better way to represent the SRC model is to express the test sample

S. Singh and A. Majumdar are with Indraprastha Institute of Information Technology, Delhi, India (e-mail: {shikhas, angshul}@iiitd.ac.in).

as a linear combination of all training samples where the weights corresponding to samples of incorrect class will be zero. This can be expressed as follows,

$$v_{test} = \sum_k V_k \alpha_k + \varepsilon = V\alpha + \varepsilon \qquad (2)$$

Here V represents all the training samples stacked as columns and α is formed by concatenating the $\alpha_k$'s vertically. The error ε is Normally distributed.

According to the SRC assumption [10], most of the coefficients in α will be zeroes. Therefore (2) is a sparse recovery problem. One can use $l_p$-norm minimization (0<p≤1) or any greedy algorithm for solving α. Once the sparse α is obtained, the task is to assign $v_{test}$ to the correct class. In SRC this is done by computing the distance between $v_{test}$ and the class representation defined by $V_k\alpha_k$. Usually a simple Euclidean distance is computed: $d_k = \|v_{test} - V_k \alpha_k\|_2$. It is expected that for the correct class this distance will be the smallest; therefore it is sensible to assign $v_{test}$ to the class having the minimum $d_k$.

This concludes the single label SRC technique. In multi-label SRC the input test sample $v_{test}$ may belong to multiple classes. Therefore instead of assigning the test sample to only one class by looking at the minimum $v_{test}$, we will consider other classes that have small $d_k$'s. For multi-label classification we can consider all classes within the range $\tau \times \min(d_k)$ to be active classes for $v_{test}$; in this work we have used τ=2. The algorithm is expressed succinctly.

ML-SRC Algorithm.

1. Solve the optimization problem expressed in (2).
2. For each class k compute class-wise distance:
   $d_k = \|v_{test} - V_k \alpha_k\|_2$
3. Assign test sample to all classes whose distance is less than $2 \times \min(d_k)$.

### III. EXPERIMENTAL EVALUATION

We have carried out experiments on two popular NILM datasets – REDD[1] and Pecan Street[2]. To emulate real-life scenario for both the datasets aggregated readings over 10 minutes have been considered. We only consider the active power as input and the data for each hour forms the length of the sample. Usually about 70~80 percent of the data is used for training the remaining for testing. Our objective is to reduce the required volume of training data, therefore in this work we consider only 10% data for training and 90% for testing. Apart from the ratio of training to test samples, the protocol remains same as [8].

The standard measures for multi-label classification based NILM have been defined in [7]. The $F1_{macro}$ and the $F1_{micro}$ are based on the popular F1 score defined for single label classification.

$$F1(TP, FP, FN) = \frac{2*TP}{2*TP + FP + FN}$$

where TP is True positive, FP is false positive and FN is false negative.

$$F1_{micro} = F1\left(\sum_{i=1}^{N} TP_i, \sum_{i=1}^{N} FP_i, \sum_{i=1}^{N} FN_i\right)$$

$$F1_{macro} = \frac{1}{N}\sum_{i=1}^{N} F1(TP_i, FP_i, FN_i)$$

Here, $TP_i$, $FP_i$ and $FN_i$ denote the number of true positives, false positive and false negative for the label i. N is the number of labels in the dataset.

These measures show how accurately an algorithm can predict the ON / OFF state of appliances. It does not prove insight into the actual energy consumption. For this purpose the second metric defined in [7] is the average energy error (AEE) defined as follows,

$$AEE = \frac{\left|\left(\sum_{i=1}^{N} Average\_power_i\right) - \left(\sum_{i=1}^{N} Actual\_power_i\right)\right|}{\left(\sum_{i=1}^{N} Actual\_power_i\right)}$$

As mentioned before [8, 9] are the most recent works on multi-label classification based NILM. Both the techniques surpass results from traditional multi-label classification algorithm like MLKNN and RAKEL. The work [8] has shown to improve over other as well as state-of-the-art deep learning techniques like multi-label stacked autoencoder. Therefore in this work we will not compare against the techniques that have already been outperformed by deep learning (DL) [8] and graph learning (GL) [9]. We also compare against the newly developing classification approach of extreme learning machine (ELM); in [11] it has been used multi-label classification. In summary we compare with three of the latest known tools in multi-label classification – DL, GL and ELM. The overall results are shown in Tables I and II.

TABLE I  RESULTS ON REDD

| Method | Macro F1-measure | Micro F1-measure | Average energy error |
|---|---|---|---|
| DL | 0.4519 | 0.4983 | 0.1433 |
| GL | 0.5662 | 0.5839 | 0.1349 |
| ELM | 0.5191 | 0.5526 | 0.8884 |
| Proposed | **0.6537** | **0.6801** | **0.0445** |

TABLE II  RESULTS ON PECAN STREET

| Method | Macro F1-measure | Micro F1-measure | Average energy error |
|---|---|---|---|
| DL | 0.6039 | 0.6049 | 0.1236 |
| GL | 0.6143 | 0.6206 | 0.1162 |
| ELM | 0.6020 | 0.6097 | 0.8989 |
| Proposed | **0.7006** | **0.7035** | **0.0338** |

We see that for the smaller REDD dataset DL produces very poor results but for Pecan Street it performs at par with the other benchmarks we have used. This is because REDD is a small dataset and 10% of the data is insufficient for DL and hence it overfits; but Pecan street is a much larger dataset and 10% of its data is sufficient for the DL to train.

---
[1] http://redd.csail.mit.edu/
[2] http://www.pecanstreet.org/category/dataport/

Note that even though ELM performs good in terms of F1 measures, the average energy error is very poor.

Of the methods compared against, GL performs the best. It performs reasonably in terms of all the metrics. But our method yields much better results than GL (and the rest), it is around 10% better in terms of all metrics.

For more granular analysis we present the appliance level results for four popular devices in Tables III and IV. As one can see the conclusions do not vary.

TABLE III
APPLIANCE LEVEL EVALUATION ON REDD DATASET

| Device | DL | | GL | | ELM | | Proposed | |
|---|---|---|---|---|---|---|---|---|
| | Error | F1-score | Error | F1-score | Error | F1-score | Error | F1-score |
| Dishwasher | 0.2902 | 0.6409 | 0.2516 | 0.6256 | 0.9667 | 0.6240 | **0.0264** | **0.7433** |
| Kitchen outlet | 0.2716 | 0.5578 | 0.3671 | 0.5071 | 0.9725 | 0.5063 | **0.0931** | **0.6631** |
| Lighting | 0.2341 | 0.5744 | 0.2907 | 0.5509 | 0.9711 | 0.5288 | **0.0373** | **0.7045** |
| Washer dryer | 0.2417 | 0.3599 | 0.3104 | 0.5390 | 0.9705 | 0.5127 | **0.0523** | **0.6990** |

TABLE IV
APPLIANCE LEVEL EVALUATION ON PECAN DATASET

| Device | DL | | GL | | ELM | | Proposed | |
|---|---|---|---|---|---|---|---|---|
| | Error | F1-score | Error | F1-score | Error | F1-score | Error | F1-score |
| Dishwasher | 0.3754 | 0.6325 | 0.1284 | 0.6369 | 0.9338 | 0.6263 | **0.0417** | **0.7569** |
| Kitchen outlet | 0.4934 | 0.5737 | 0.1957 | 0.5531 | 0.9718 | 0.5429 | **0.1620** | **0.6715** |
| Lighting | 0.3163 | 0.6129 | 0.1184 | 0.6145 | 0.9466 | 0.6252 | **0.0679** | **0.7198** |
| Washer dryer | 0.3384 | 0.5996 | 0.1659 | 0.6017 | 0.9640 | 0.5578 | **0.0734** | **0.7037** |

IV. CONCLUSION

In recent times, there is a concerted effort towards truly non-intrusive load monitoring [7-9, 12]. This is required for practical large scale roll-out of NILM as a service with the larger goal of improving energy sustainability. In this respect the multi-label classification framework has been showing promise [7-9]. However, recent deep learning based solutions like [8] require large volume of labeled training data. In order to reduce that requirement we propose a simple solution based on adapting the SRC framework [10] to solve multi-label classification problems. Our proposed multi-label SRC improves over state-of-the-art techniques by a considerable margin.

The main shortcoming of our approach is that, it is not possible to estimate different stages of each appliance via our method. Neural network based function approximation approaches may be better in this respect.

SRC is the basic algorithm. Over the years various modification have been proposed, such as kernel SRC [13] group SRC [14], dictionary learnt SRC [15] etc. We plan to adapt all the popular variants to solve multi-label classification problems.